\documentstyle[aps,epsf]{revtex}

\tighten
\begin{document}
\draft

\title{Strongly Localized State of a Photon at the Intersection of
the Phase Slips in 2D Photonic Crystal with Low Contrast of 
Dielectric Constant
}

\author{V. M. Apalkov and  M. E. Raikh}
\address{Department of Physics, University of Utah, Salt Lake City,
UT 84112, USA}
\maketitle

\begin{abstract}
Two-dimensional photonic crystal with a rectangular symmetry and 
low contrast ($< 1$) of the dielectric constant is considered. 
We demonstrate
that, despite the {\em absence} of a    
bandgap, strong localization of a photon can be achieved
for certain ``magic'' geometries of a unit cell 
 by introducing two $\pi/2$ phase slips along the major axes. 
Long-living photon mode is bound to the intersection of the phase slips.
We calculate analytically the lifetime of this mode 
for the simplest geometry
-- a square lattice of cylinders  
of a radius, $r$. 
We find the magic radius,  $r_c$, of a cylinder to be $43.10$
percent of the lattice constant. 
For this value of $r$, the quality factor
of the bound mode exceeds $10^6$. Small ($\sim 1\%$) 
deviation of $r$ from $r_c$
results in a drastic damping of the bound mode.  
\end{abstract}

\vspace{4mm}

{\em Introduction}.--The field of  photonic bandgap materials 
was pioneered by two papers\cite{yablonovich,john87}.
In Ref. \cite{yablonovich} the idea of bandgap-induced inhibition 
of spontaneous emission, which limits the  
 performance of light-emitting devices, was put forward. In 
Ref. \cite{john87} the observation was made
that a photonic bandgap facilitates  strong Anderson localization 
of light, which otherwise is hard to achieve\cite{genack01}. 
 Both papers addressed the issue of defect-induced 
localized photonic modes.
However, the  defects considered in Refs. \cite{yablonovich} 
and \cite{john87} 
were of a  different nature. In  \cite{john87}
the defects were  point-like scatterers, i.e. {\em local} perturbations
of the {\em spatially periodic} dielectric constant. 
In contrast,
a defect discussed in  \cite{yablonovich}
represented a  {\em periodicity interruption}  or, in other
words,  {\em  a phase slip}\cite{mccal85,avrutin88}. 
While both types of defects  cause the formation of the in-gap
states, there is a fundamental difference between them.  
A point-like defect does not lift the long-range order of the underlying
crystal.  On the contrary, in the presence of a phase-slip, the
distance between two lattice sites located to the left and to the
right from the phase slip is non-integer (in the units of the lattice
constant).

Lately, the properties of defects in photonic bandgap structures
(phase-slip-like\cite{foresi97} and more complex defects, specifically 
designed for certain  manipulations of the light flow\cite{joan97})
have become a branch of research of their own\cite{joan02}.

 Theoretical studies of photonic bandgap
materials are almost exclusively  focused on  structures
with a high contrast of the dielectric constant which
is sufficient 
for the formation of either 2D or 3D  
bandgap, i.e. the region
of frequencies where the propagation of light is completely 
forbidden.  
For low contrast, periodicity manifests itself
in the formation of  narrow stop bands only along certain directions
close to the Bragg condition. The overall density of photonic 
states is weakly perturbed by such a periodicity. 
This rules out the  
possibility of defect-induced localized modes.
Although, strictly speaking, this statement is 
correct, we demonstrate in the present paper that strongly
localized photon modes are possible in photonic crystals with
contrast of dielectric constant $< 1$. 
By ``strongly localized'' modes we mean the modes with
decay time much longer than the inverse frequency, or, in other words,
with high quality factor, $Q$. In fact,  we will demonstrate below
that, when the contrast of the dielectric constant 
is low, the proper choice of the geometry and parameters 
of the 2D unit cell, 
combined with proper periodicity interruptions, allows to achieve
a $Q$-factor of the localized mode as high as $Q\sim 10^6$.

{\em Qualitative Consideration}.-- In the absence of photonic gap,
it can be concluded on general grounds that formation of a
defect-induced localized mode is impossible, since its frequency would
be degenerate with the continuum of the extended Bloch waves.  There is,
however, an exception. Consider a 2D photonic crystal in which
dielectric constant $\epsilon(x,y)$ is {\em separable}, $\epsilon(x,y)
= \epsilon_0+\epsilon_1(x) + \epsilon_2(y)$, where $\epsilon_1$, and
$\epsilon_2$ are weak periodic modulations with periods $a$ and $b$,
respectively. 
Thus, for the light wave propagating
strictly along the $x$-direction the position of a narrow stop band
in the
dispersion law is determined by the Bragg condition
$\epsilon_0^{1/2}\omega_x = \pi c/a$. 
Analogously, for the light wave along the $y$-direction, 
the Bragg condition reads 
$\epsilon_0^{1/2}\omega_y = \pi c/b$. Assume now, that the phase slips 
along both directions, $x$ and
$y$, are introduced into the photonic crystal.  This situation is
illustrated in Fig. 1. Note, that, even in the presence of the 
phase slips,
$\epsilon(x,y)$ remains separable. On the other hand, it is known that,
much like a point defect, a phase slip in one dimension
causes a localized photonic state that decays exponentially 
away from the phase slip\cite{mccal85}. 
Then, in the presence of two phase slips, a {\em truly} 
localized photonic mode bound to the intersection of the phase 
slips, $x=y=0$, exists. The frequency of this mode is close 
to $\omega_0 = \left(\omega_x^2 + \omega_y^2\right)^{1/2}$.  
We would like to emphasize that existence
of a 2D localized mode, degenerate with continuum, is unique for the
case of the phase slips, since this mode is effectively {\em decoupled}
from the continuum. On the other hand, even for separable
$\epsilon(x,y)$, a {\em point defect} of arbitrary strength is unable
to localize a photon when the gap is absent.
This is because, in
the presence of a defect, the separability is lifted.  

Naturally, any
violation of separability of $\epsilon(x,y)$ would result in the
leakage of a mode bound to the intersection of the phase slips. 
It is also obvoius that separability is lacking for any  realistic 
geometry of photonic
crytal (like the one illustrated in Fig.~1).  
Our prime
observation is that, for certain ``magic'' geometries of the unit
cell, the leakage can be suppressed even without separability.
Then  the lifetime of the
mode remains much larger than $\omega_0^{-1}$. 

We start with the remark that, unlike the case of a random disorder, 
the decay of the localized mode is, to the first approximation, 
determined by a {\em single} Fourier component of $\epsilon(x,y)$. 
Our main idea is that this component  {\em can be eliminated} by a
proper arrangement of a unit cell.
As the simplest example, consider a rectangular
lattice  of cylinders of radius $r$ (Fig.~1a). 
In this case, the Fourier component, 
responsible for the leakage, is proportional to $J_1 
\left( 2\pi \sqrt{ (r/a)^2 +(r/b)^2} \right)$, 
where $J_1$ is the Bessel function of the first order. 
Correspondingly, the magic value of 
$r$, for which the leakage is suppressed, is determined by a first zero 
of $J_1(u)$, i.e. $u=u_0\approx 3.830$. This yields for the magic 
radius, $r_c$, the value  $r_c = 0.610 ab/\sqrt{a^2+b^2}$. 
In particular, for a 
square lattice, $a=b$,  we obtain $r_c = 0.4310 a$.
Away from the magic radius, the quality   
factor falls off as $(r-r_c)^{-2}$,
\begin{equation}
Q^{-1} = 
\frac{\mbox{Im} \omega }{\omega }= 
 C \left( \frac{\delta \epsilon }{\epsilon _0} \right)
                   \left( \frac{r-r_c }{a} \right)^2,
\label{Q1}
\end{equation}
where $\delta \epsilon $ is the difference between 
dielectric constants of the cylinder and the background. 
For the coefficient $C$ in the most interesting case of 
$\pi/2$ phase slips we obtain below 
\begin{equation}
C = \left(\frac{2^{11/2}}{15}\right) 
  \frac{u_0 J_0^2 (u_0)}{J_1(2^{-1/2}u_0)}
     \approx 4.3 ,
\end{equation}
where $J_0 $ is the Bessel function of a zero order. 
In the limit $r \rightarrow r_c$ the growth of $Q$ saturates
due to the higher-order processes illustrated in Fig.~2. 
 These processes involve, strictly speaking, all the
Fourier components with non-zero 
$n$ and $m$; these components are proportional to 
$ J_1 \left( 2\pi \sqrt{ (nr/a)^2 +(mr/b)^2} \right)$.
Our important observation is that the leakage can be 
further suppressed by 
 {\em fine tuning} of $r$ around $r_c$.
 This is due to the
compensation of  different contributions -- phenomenon analogous to
the Fano resonance\cite{fano61}. The 
final expression for the 
quality factor, which incorporates the higher-order 
processes, has the form
\begin{equation}
Q^{-1} =  C \left( \frac{\delta \epsilon }{\epsilon _0} \right)
    \left[  \left( \frac{r-r_c}{a} - 
          \frac{\delta \epsilon }{\epsilon _0} 
\kappa _1 \right)^2 + 
\left( \frac{\delta \epsilon }{\epsilon _0} \right)^2 \kappa_2 ^2
\right] ,
\label{Q2}
\end{equation}
where dimensionless factors $\kappa _1$ and $\kappa _2$ 
involve the following lattice sums 
\begin{equation}
\kappa _2  = \left(\frac{3^{1/2}}{21 \pi^4}\right) 
   \frac{a^4}{u_0 J_0(u_0)} 
\sum _{n,m>0}
\frac{ F_{n,m}^2 + F_{n+1,m+1}^2 + 2 F_{n,m+1}^2 }
{n(n+1) + m(m+1)}     ,  \label{S2}
\end{equation}
\begin{equation}
\kappa _1 = 
 \left( \frac{3}{2}\right)^{3/2} \!\!\! \kappa _2 + \frac{a^4}{2^{1/2} 
\pi ^4 u_0 J_0(u_0 )}
      \sum _{n,m>0}
\frac{ F_{n,m}F_{n+1,m+1}}{n(n+1) + m(m+1)}.  
                  \label{S1}
\end{equation}
 The elements $F_{n,m}$ are related
to the Fourier 
components of $\epsilon (x,y)$ as follows 
\begin{equation}
F_{n,m} = 
  \frac{\pi u_0}{2^{1/2}(n^2+m^2)^{1/2}a^2} ~~
J_1  \! \left( 2^{-1/2} (n^2+m^2)^{1/2} u_0 
            \right) .
\label{Fnm}
\end{equation}
The sum in Eq.~(\ref{S2}) reflects the processes $2-2_1$ in Fig.~2,
while the sum in Eq.~(\ref{S1}) originates from the process 
$2-2_2$.  
First term in Eq.~(\ref{Q2}) describes the Fano resonance. 
It suggests that the leakage can be reduced, if $r$ slightly exceeds 
$r_c$ by $(\delta \epsilon /\epsilon _0) \kappa _1 a$. Then
the maximal possible value of the quality factor is given by 
\begin{equation}
Q _{m} =  \frac{1}{C\kappa _2^2} \left( \frac{\epsilon _0 }{
 \delta \epsilon }\right)^3 ,
\label{Q3}
\end{equation}
Formula (\ref{Q3}) is our main quantitative result. In general, 
it could be expected that parameters $\kappa _1$ and $\kappa _2$ are 
$\sim 1$. Remarkably, 
their values  turn out to be very small,  
$\kappa _1 \approx 5 \cdot 10^{-3} $ and $\kappa_2 
 \approx 0.8 \cdot 10^{-3} $.
Thus, even for $\delta \epsilon =\epsilon _0 $, we get a very high 
quality factor, $Q_{m} \approx  0.4 \cdot 10^{6}$.
Achieving this value, however, requires 
a high precision in the choice of $r$.
For example, without fine tuning 
(i.e. for $r=r_c$), Eq.~(\ref{Q2}) suggets that the quality factor 
drops from $Q_m$ to 
$\frac{\kappa _2^2}{\kappa _1^2 +\kappa_2^2} Q_m \approx 0.03 Q_m$.

{\em Derivation of Eqs.~(\ref{Q1}), (\ref{Q2})}.- 
For simplicity we consider the case of TM polarization with
electric field, ${\cal E}(x,y)$, along the $z-axis$. 
Since the derivation is based on the perturbation theory, 
it is convenient to cast the wave equation for ${\cal E}(x,y)$ into
Schr\"odinger-like form
\begin{equation}
\left( \frac{ \partial^2}{\partial x^2} + U_1 (x) \right) {\cal E} +
 \left( \frac{ \partial^2}{\partial y^2} + U_2 (y) \right) {\cal E} +
U^{(ps)}_{pert}(x,y) {\cal E}= - \lambda {\cal E}   \label{eq1},
\end{equation}
where $\lambda = \frac{\omega ^2}{c^2} \epsilon _0$ stands for the 
``energy'', while the ``potentials'' $U_1(x)$, $U_2(y)$, and  
$U^{(ps)}_{pert}(x,y)$ are defined as 
\begin{equation}
U_1 (x) =\sum _n U_{n,0}(x + d _x \mbox{sign} (x)), 
~~~~U_2 (y) = \sum _m U_{0,m}(y + d _y \mbox{sign} (y)), 
\label{U1}
\end{equation}
\begin{equation}
U^{(ps)}_{pert} (x,y)  = \sum _{n,m > 0}
U_{n,m}^{(ps)}(x,y) = \sum _{n,m > 0}
  U_{n,m} (x+ d _x \mbox{sign} (x),y+ d_y \mbox{sign} (y)).
\label{U_p}
\end{equation}
The arguments in r.h.s. of Eqs.~(\ref{U1}), (\ref{U_p}) 
reflect the periodicity interruptions by $2d_x$ and $2d_y$, shown
in Fig.~1(a), while the functions $U_{n,m}(x,y)$ are the components   
of the Fourier expansion of {\em periodic} $\epsilon (x,y)$ in 
the absence of the phase slips
\begin{equation}
\frac{\omega ^2 }{c^2} \epsilon (x,y) 
= \sum _{n,m \geq 0} U_{n,m}(x,y) =
\left(\frac{\delta \epsilon }{\epsilon _0} \right)
   \sum_{n,m \geq 0 } \left( F_{n,m} e^{2i( ng_x x+ m g_y y)} + 
h.c. \right),
\end{equation}
where $g_x = \pi /a $ and $g_y = \pi /b$.
In these notations, the known solution\cite{mccal85,avrutin88} 
for the localized mode induced by a one-dimensional 
phase slip takes the form 
\begin{equation}
 E^{(x)}_0(x)=(2\gamma _x )^{1/2}
    \cos (g_x x) \exp\left(-\gamma _x |x|\right), ~~~~
 E^{(y)}_0(y)=(2\gamma _y )^{1/2}
  \cos (g_y y) \exp\left(-\gamma _y |y|\right) ,
\end{equation}
where the decrements $\gamma _x$, $\gamma _y$ are given by  
\begin{equation}
\gamma _x = \left(\frac{\delta \epsilon }{\epsilon _0} \right) 
 \left|  \frac{F_{1,0}}{2 g_x}  \sin \left( \frac{2\pi d _x}{a}\right) 
    \right| ,
~~~~
\gamma _y = \left(\frac{\delta \epsilon }{\epsilon _0} \right) 
  \left| \frac{F_{0,1}}{2 g_y} \sin \left( \frac{2\pi d _y}{b}\right)
 \right|.
\end{equation}
Therefore, the zero-order solution of Eq.~(\ref{eq1}) corresponding to
$U_{pert}^{(ps)} \equiv 0$ is 
$
{\cal E}_0 (x,y) =  E^{(x)}_0(x) 
  E^{(y)}_0(y)$ with the ``energy''
\begin{equation}
\lambda _0 = g_x^2+g_y^2 +
   \delta \epsilon 
\left[ \left| F_{1,0} \right|  
  \cos \left( \frac{2\pi d _x }{a} \right)
           +  \left| F_{0,1} \right| 
       \cos\left( \frac{2\pi d _y }{b} \right)
\right]   .
\label{lambda0}
\end{equation}
The contour of equal intensity for ${\cal E}(x,y)$ is shown 
schematically in Fig.~1b.

Calculation of leakage of the localized mode ${\cal E}_0 (x,y)$
requires taking into account the phase-slip-induced 
perturbation, $U_{pert}^{(ps)}(x,y)$, 
up to the second order. We will restrict further consideration to the 
most interesting case of $\pi /2$ phase slip, i.e. $d_x = a/4$, 
$d_y = b/4$. 
For this purpose we will need, in addition to
 ${\cal E}_0 (x,y)$, the 
zero-order solutions of Eq.~(\ref{eq1}), corresponding to the
continuous spectrum of propagating modes. These solutions 
${\cal E}^{\mu, \nu}_{p, q} (x,y)= E^{(x)}_{\mu, p}(x)
         E^{(y)}_{\nu ,  q} (y) $ 
can be easily found in a similar way as  ${\cal E}_0 (x,y)$
\begin{equation}
 E ^{(x)}_{\mu , p}(x)  = \frac{1}{\sqrt{
 (1+ \tilde{p}^2)}}
     \left\{ e^{-i g_x x} |\sin(p x)|
   - e^{i g_x x} \left[ \tilde{p} \cos (p x) +
      i \mu  (1+\tilde{p}^2)^{1/2} \sin (px)  \right]
 \right\},
 \label{deloc1}
\end{equation}
where $\tilde{p} = 2 p g_x \epsilon _0 /(\delta \epsilon 
\left| F_{1,0}\right| )$, and
the index $\mu \pm 1$ enumerates the upper and the lower branches 
of a one-dimensional spectrum in the $x$--direction; these branches are 
 separated by a narrow Bragg gap 
$\delta \epsilon \left| F_{1,0}\right|$ 
(see Fig.~2). 
The ``energies''
corresponding to ${\cal E}^{\mu, \nu}_{p, q} (x,y) $ are given by 
\begin{equation}
\lambda ^{\mu , \nu }_{p, q}  = g_x^2+g_y^2 + \delta \epsilon 
\left[  \mu 
\left| F_{1,0} \right| \left( 1 + \tilde{p}^2 \right)^{1/2}
+ \nu  
\left| F_{0,1} \right| 
   \left( 1 + \tilde{q}^2 \right)^{1/2}   \right]    .
\label{deloc2}
\end{equation}
The form of the wave equation (\ref{eq1}) allows to write down 
automatically the epression for the first and second-order 
corrections, $\lambda _0^{(1)}$, $\lambda _0 ^{(2)}$, 
to the ``energy'' $\lambda _0$, Eq.~(\ref{lambda0}). 

\begin{equation}
\lambda _0^{(1)} =  \langle {\cal E}_0 |U^{(ps)}_{pert}|
     {\cal E}_0 \rangle  , ~~~
\lambda _0^{(2)} =  \sum _{\mu , p, \nu , q} \frac{  |\langle {\cal E}_0 | \hat{T}_{1,2}
          |  {\cal E}^{\mu , \nu }_{p, q} \rangle |^2 }{ \lambda _0 -
        \lambda ^{\mu , \nu }_{p , q}} ,
\end{equation}
Obviously,  $\lambda _0^{(1)}$ does not cause any leakage, and,
moreover, $\lambda _0^{(1)}\equiv 0$ due to symmetry of the geometry
under consideration. 
First non-vanishing contribution to  $\mbox{Im} \lambda $, which 
describes the leakage, as follows from the golden rule,
\begin{equation}
\mbox{Im} \lambda  =  \mbox{Im} \lambda _0^{(2)}=
   \pi    \sum _{
        \mu , p,  \nu , q}
        |\langle {\cal E}_0 | \hat{T}_{1,2}
          |  {\cal E} ^{\mu , \nu }_{ p, q} \rangle |^2 \delta ( \lambda _0 -
        \lambda ^{\mu \nu }_{p,q} ) ,
\label{im1}
\end{equation}
emerges if the lowest order expression for the operator 
$\hat{T}_{1,2}$ 
\begin{equation}
 \hat{T}_{1,2}^{(0)} = U_{1,2}^{(ps)} (x,y) + U_{2,1}^{(ps)} (x,y)
\end{equation}
is substituted into Eq.~(\ref{im1}).
Evaluation of $\mbox{Im} \lambda $ leads to Eq.~(\ref{Q1}) for 
the quality factor, $Q= 2\lambda /\mbox{Im} \lambda =
\omega /\mbox{Im} \omega $. Since for magic configuration 
this term is zero, the operator 
$\hat{T}_{1,2}$ should be taken with higher accuracy.
Namely, higher order processes, illustrated in Fig.~2, amount to the 
following modification of the operator $\hat{T}_{1,2}$
\begin{equation}
\hat{T}_{1,2} =  \hat{T}_{1,2}^{(0)} +
\sum _ {m,n>0}
\sum _ {m_1,n_1>0}
 \sum _ {p_x, p_y}
 \frac{ U_{m,n}^{(ps)} | {\cal E}_{p_x, p_y} \rangle \langle {\cal E}_{p_x, p_y}
   | U_{m_1,n_1}^{(ps)} }{\lambda _0 - \lambda _{p_x,p_y}}
    \left( 1 - \delta_{m,1} \delta_{n,1} 
   \delta_{m_1,1} \delta_{n_1,1} \right),
\label{t12}
\end{equation}
where ${\cal E}_{p_x, p_y} = \exp (ip_x x + ip_y y)$, 
$\lambda _{p_x,p_y} = p_x^2+p_y^2$ stand for
 ``nonresonant'' solutions of Eq.~(\ref{eq1}) with $|\bbox{p}|> 
2\pi /a$. For these solutions, periodic modulation can be neglected.
Matrix elements of $U_{n,m}^{(ps)}$ between ${\cal E}_0 (x,y)$ and
plane waves, required for evaluation of Eq.~(\ref{im1}),  
can be calculated analytically. 
The form of these matrix elements suggests that the main 
contribution to 
$\mbox{Im} \lambda $ comes from the small angular regions $\sim \delta 
\epsilon/\epsilon _0$ in Fig.~2 (of the order of the Bragg gap)
around the Bragg directions.  
After that, the summation in 
Eq.~(\ref{im1}) with the use of the $\delta $-function, reduces to the 
integrals    
\begin{equation}
I_n = 
    \int_0 ^{\infty } du ~ u (1+u)^{-(2n+5)/2} = \frac{4}{(2n+1)(2n+3)}
\end{equation}
with $n=1,2,$ and 3. Now, as summation over all momenta in 
Eqs.~(\ref{im1}), (\ref{t12}) is 
performed analytically, we arrive at Eq.~(\ref{Q2}).

{\em Conclusion}.- First photonic-crystal based lasers were
reported in Refs. \cite{panter99,meier99}. In Ref. \cite{panter99}
a two-dimensional photonic crystal was fabricated within
the active region of the {\em InGaAsP} multiple quantum well
structure using the etching procedure. The contrast of the dielectric
constants between the semiconductor and etched air-holes was 
high ($\approx 10:1$). This contrast was sufficient for 
opening of the two-dimensional bandgap. Then a defect,
which was essentially a missing air-hole\cite{panter99},  
gave rise to the in-gap mode
localized within approximately two periods of the host hexagonal
crystal.  It was demonstrated in Ref. \cite{panter99} that above the
threshold the laser emission is dominated by this defect mode.

A low-contrast photonic crystal with array of holes etched 
in the $SiO_2$ substrate ($\epsilon=1.46$) of a solid organic 
gain film ($\epsilon=1.7$) was reported in Ref.\cite{meier99}. 
Without photonic gap, the  authors did not attempt to create 
a localized mode, although in the later work\cite{meier'99} a 
square  symmetry of the lattice pattern was employed. 

In the present paper we have demonstrated that a localized photon 
mode is possible in a 2D photonic crystal with a {\em low contrast}
of the dielectric constant. In a sense, the absence of photonic 
gap can be compensated by a high-precision tailoring of the 
{\em unit cell } parameters. 
There is another example, when an accurate choice of {\em point-defect} 
parameters gives rise to a long lifetime. It pertains to the crystals with
a {\em high contrast} of the dielectric 
constant\cite{painter99',johnson01}. Namely, in a finite thickness
2D photonic crystal with a complete gap for in-plane light 
propagation, the leakage of the in-gap mode in the $z$--direction 
can be strongly suppressed for a certain radius of a cylinder
constituting a defect\cite{johnson01}.

A natural extension of the present study is to explore a possibility
to guide light from one location to another using an {\em array} of 
phase silps in low-contrast magic crystals. 
It would be appealing if such  
crystals  offered an alternative to the conventional
bandgap materials\cite{joan02} in terms of manipulating the light flow.

This work was supported by the Army Research Office under Grant
No. DAAD 19-0010406 and the Petroleum Research Fund under Grant
No. 37890-AC6.

\end{document}